\documentclass[aps,prd,twocolumn,superscriptaddress,nofootinbib,floatfix]{revtex4-1}
\usepackage{graphicx,bm,dcolumn,color,epsfig,amsmath,amssymb,rotating,float,comment}
\newcommand{\ba}{\begin{eqnarray}}
\newcommand{\ea}{\end{eqnarray}}

\newcommand{\D}{\nabla}

\newcommand{\4}{\frac{1}{4}}

\newcommand{\Fc}{\mathcal{F}}

\newcommand{\la}{\lambda}

\newcommand{\sa}{\sigma}

\begin{document}

\title{UV completion of the Starobinsky model, tensor-to-scalar ratio, and constraints on non-locality}
\author{James Edholm}
\affiliation{Consortium for Fundamental Physics, Lancaster University, Lancaster, LA1 4YB, United Kingdom}

\begin{abstract}
In this paper, we build upon the successes of the ultraviolet (UV) completion of the Starobinsky 
model of inflation. This involves an extension of the Einstein-Hilbert term by an {\it infinite covariant derivative} theory of gravity which is quadratic in curvature. It has been shown that such a 
theory can potentially resolve the cosmological singularity for a flat, homogeneous and 
isotropic geometry, and now it can also provide a successful cosmological 
inflation model, which in the infrared regime matches all the predictions of the Starobinsky model of inflation.
The aim of this note is to show that the tensor-to-scalar ratio is modified by the scale of non-locality, and in 
general a wider range of  tensor-to-scalar ratios can be obtained in this class of model, which can put
a lower bound on the scale of non-locality for the first time as large as the ${\cal O}(10^{14})$~GeV.
\end{abstract}
\maketitle
\section{Introduction}

Einstein's theory of General Relativity (GR) has been extremely successful in its predictions about the infra-red (IR) regime~\cite{Will}. 
However, in the ultraviolet (UV) regime, the theory exhibits pathologies,  and at a quantum level the theory becomes 
non-renormalizable. Classically, GR allows both black hole and cosmological type singularities.
It has been known for some time from a seminal paper by Stelle~\cite{Stelle:1976gc} that quadratic curvature gravity is renormalizable, but 
 in general it suffers from the presence of a Weyl {\it ghost}~\footnote{This is a generic problem for any higher derivative theory, i.e. 
 more than 2 derivatives, where extra derivatives count for extra poles in the propagator and extra degrees of freedom other than the 
 original degrees of freedom. The extra poles generically harbour ghosts~\cite{Van}. }.

In \cite{Biswas:2005qr} and \cite{Biswas:2011ar}, it was demonstrated that one can tame the problem of ghosts in a quadratic curvature gravity, provided one invokes {\it infinite covariant derivatives} acting on the curvature~\footnote{In \cite{Biswas:2011ar}, the authors constructed the most generic quadratic curvature gravity involving Ricci scalar and tensor, and the Riemann/Weyl, which is ghost free and singularity free around Minkowski background.}. In this case, the graviton propagator is modified by these infinite derivatives, but it is still possible to retain the original massless graviton degrees of freedom without introducing any new poles  in either the spin-0 or spin-2 component of the graviton propagator. This can be achieved if the propagator is modified by an exponent of an {\it entire function}, i.e. $e^{\gamma(\Box)}$, where $\gamma(\Box)$ is an entire function of the d'Alembertian, $\Box=g^{\mu\nu}\nabla_{\mu}\nabla_{\nu}$, where
the Greek indices run from $0,1,2,3$. The exponential of an entire function contains no roots by construction. If $\gamma > 0$, and $\Box \rightarrow \infty$,
the propagator is even more convergent than a polynomial of finite degree, thus improving upon the UV properties~\cite{Tomboulis,Modesto,Talaganis:2014ida,Talaganis:2016ovm}. In the IR, one recovers the original graviton propagator~\cite{Biswas:2011ar,Biswas:2013kla}. 

The quantum UV aspects are also improved,   due to the fact that the vertex interactions in such  infinite derivative theories, with exponential modification of the propagator in the UV, become non-local.
The scale of these modifications is governed by the scale of non-locality $M$. This has been illustrated in thermal aspects of string theory~\cite{Biswas:2009nx}, and in an improved higher derivative extension of the electroweak Standard Model~\cite{Biswas:2014yia}. It has also been shown that ultra high energy trans-Planckian scatterings in this case do not blow up in the UV~\cite{Talaganis:2016ovm}.  

From the classical point of view, in the linear regime the theory has resolved the Newtonian singularity,
and the blackhole singularity for mini blackholes~\cite{Biswas:2011ar}~\footnote{See also~\cite{Tseytlin,Siegel:2003vt},
where infinite derivative corrections to curvature have been proposed from string theory and string theory.  One would 
expect infinite higher derivative corrections in the gravitational sector, due to $\alpha'$ corrections.}, 
and also the dynamical formation of such blackholes, which have no horizon and no Schwarzschild's singularity~\cite{Frolov}.

The initial motivation of studying {\it ghost-free} infinite derivative gravity was to resolve the cosmological Big Bang singularity problem 
in Einstein's theory of gravity, by supplanting it with a Big Bounce~\cite{Biswas:2005qr}, 
and to study cosmological perturbations around the Big Bounce~\cite{Conroy:2014dja}. In  
\cite{Biswas:2010zk,Biswas:2012bp}, it was shown that sub and super Hubble perturbations around the bouncing solution are stable. 
It was already pointed out in \cite{Biswas:2005qr,Biswas:2012bp,Biswas:2013dry,Chialva:2014rla,Koshelev:2016vhi} that such a quadratic action of gravity 
would serve as a UV completion of the original Starobinsky model of inflation~\cite{Starobinsky:1980te}.
In fact, \cite{Biswas:2013dry} already mentioned the possibility of explaining low multipoles 
observed in the temperature anisotropy of the cosmic microwave background radiation (CMBR)~\cite{WMAP,Ade:2015xua}, and its connection with a bouncing cosmology in connection with a UV-improved Starobinsky inflation. 
A rigorous proof of avoiding the cosmological singularity was provided in \cite{Conroy:2016sac}.

Indeed, the scale of non-locality $M$ is one of the key parameters for any such ghost-free higher derivative modification of gravity.
It is important to constrain this parameter from all possible observations. One of the best constraints on $M$ arises from the fact that at table-top
experiments it is possible to constrain the departure from the $1/r$-fall of Newtonian gravity~\cite{Kapner:2006si}, which has been tested up to 
$5\times 10^{-6}$ meters, and this places a constraint on the scale of non-locality of $M > 0.004$~eV~\cite{Edholm:2016hbt}.
This is a very weak but useful constraint. 

The main aim of this paper is to improve the constraint on the scale of non-locality, i.e.~$M$, from cosmological observations, such as inflationary cosmological perturbations~\cite{Mukhanov:1990me}. 
 
Primordial inflation~\cite{Guth,Linde,Albrecht} is currently one of the best paradigms to explain the temperature anisotropy in the CMBR and large scale structures in the Universe. It can occur in many 
different sectors, such as in the visible sector~\cite{La,Berzukov,Allahverdi:2006iq}, individually or simultaneously~\cite{Liddle:1998jc}, and for a review one can read~\cite{Mazumdar:2010sa}. 
However, inflation in the gravitational sector is perhaps one of the most natural ways to describe the Universe, first envisaged by Starobinsky~\cite{Starobinsky:1980te}.
The original model was described by a quadratic curvature action of gravity, and now that we have this infinite derivative modification, which improves the UV aspects of gravity, 
we should revisit its cosmological properties~\footnote{ In fact, \cite{Salvio:2015kka} highlights the robustness of Starobinsky inflation from the quantum corrections point of view for physics beyond the Standard model.~\cite{Briscese:2012ys} showed that under a suitable truncation, infinite derivative gravity can give rise to Starobinsky inflation.}.

Recently, in \cite{Biswas:2016etb,Biswas:2016egy} the authors constructed the most general quadratic curvature, infinite derivative theory of gravity which is free from ghosts 
and instability around de Sitter and anti-de Sitter backgrounds~\footnote{For parity invariant and torsion free gravity}. In order to understand the stability of the action around  de Sitter and Anti-de Sitter backgrounds, the authors of \cite{Biswas:2016egy} expanded 
the action up to second order in scalar, vector and tensor modes. It was shown that the only propagating modes would be the scalar and tensor modes. 

With the help of these mathematical tools,
in~\cite{Craps:2014wga} and in \cite{Koshelev:2016xqb}, the authors investigated the scalar and tensor perturbations in an inflationary background for infinite derivative theory of gravity. 
It was found that in the low energy limit, the scalar and tensor perturbations evolve in exactly the same way as in the Starobinsky model of inflation, 
but in the UV regime there are some subtle differences, which were highlighted in~\cite{Koshelev:2016xqb}, and we will briefly review them here. 
We will use the latest bounds on tensor modes to constrain the value of non-locality $M$. 
We will also explicitly compute the spectral tilt for the scalar perturbations for the IDG model of inflation for the first time.

\section{Infinite derivative gravity }

The most generic quadratic curvature gravity in $4$ dimensions,  which can be made {\it ghost-free} can be written in terms of the Ricci-scalar, $R$, the symmetric traceless tensor, 
$S_{\mu\nu}=R_{\mu\nu}-\frac{1}{4}Rg_{\mu\nu}$, analog with the Einstein tensor, $R_{\mu\nu}$ (the Ricci tensor), and the Weyl tensor: $C_{\mu\nu\alpha\beta}$. The $S$-tensor vanishes on maximally symmetric backgrounds~\cite{Biswas:2016etb}~\footnote{The original action was written in terms of $R_{\mu\nu}$ and 
$R_{\mu\nu\lambda\sigma}$ in~\cite{Biswas:2011ar}. However there is no loss of generality in expressing the action as~Eq.~(\ref{ligoproperaction})~\cite{Biswas:2016etb}.}: 
\begin{eqnarray}
S = \int d^4x \sqrt{-g}\left[\frac{M_P^2}{2} R
+\frac{\lambda}2\Bigg(R\Fc_1(\Box)R \right.\nonumber \\
\left. +S_{\mu\nu}\Fc_2(\Box)S^{\mu\nu}+ 
C_{\mu\nu\la\sa}\Fc_{3}(\Box)C^{\mu\nu\la\sa}\Bigg)\right]\,,
\label{ligoproperaction}
\end{eqnarray}
where Greek indices $\mu$, $\nu$ etc. run from 0 to 3, $M_P^2$ is the Planck mass, and $\lambda$ is a dimensional coupling accounting for the higher curvature modification, 
and the $\Fc_i$ are Taylor expandable (i.e. analytic) functions of the covariant d'Alembertian~\cite{Biswas:2011ar}, i.e.
\begin{equation}
       \Fc_i(\Box)=\sum_{n=0}^\infty c_{i_n}\Box^n/M^{2n}\,,
        \label{scale}
\end{equation}
where $M$ is the scale of non-locality and $c_{i_n}$ are the coefficients of the series. Using the fact that the $1/r$ fall of the Newtonian potential continues until 
around $5\times 10^{-6}$m~\cite{Kapner:2006si}, we can say that $M>10^{-2}$eV~\cite{Edholm:2016hbt}, a reasonably weak constraint. Previous work  on non-local theory, 
without the Weyl term in Eq.~(\ref{eq:occurenceunreducedaction}), used inflation data to estimate
 $M$ in that simplified version of the theory as  $M\sim$ $10^{15}$~GeV~\cite{Biswas:2013dry} 
and $M>10^8$~GeV~\cite{Biswas:2006bs}.

In fact, it was already shown in \cite{Biswas:2011ar}, that one can switch off either ${\cal F}_{i}(\Box)$ in~Eq.~(\ref{ligoproperaction}) without loss of generality, i.e. 
without introducing ghosts in the spectrum and without modifying the graviton propagator in the IR. The full equations of motion for the action have been derived in \cite{Biswas:2013cha}.

For the purpose of investigating inflation, one can use the ``redundant functions" method to set $F_2(\Box)=0$ and therefore study the following action without loss of generality~\cite{Conroy:2016sac}  
\ba \label{eq:occurenceunreducedaction}
           S = \frac{1}{2} \int d^4 x \sqrt{-g} \Big[ M^2_p R + \lambda\Big( R {\cal F}_1 (\Box) R \nonumber\\
           + C_{\mu\nu\rho\sigma} {\cal F}_3(\Box) C^{\mu\nu\rho\sigma} \Big) \Big],
\ea           
In order to solve the infinite covariant derivatives, one can consider a simple ansatz 
\ba \label{eq:boxrequalsransatz}
        \Box R = r_1R
\ea
where $r_1$ is a constant, which produces the relation ${\cal F}_1(\Box)R = F_1 R$, where $F_1$ is a constant.~\footnote{The more general ansatz $\Box R= r_1 R + r_2$, where $r_1,~r_2$ are constants, 
was used originally in~\cite{Biswas:2005qr} and then~\cite{Biswas:2010zk,Deser:2013uya,Biswas:2012bp,Craps:2014wga}. 
Setting $r_2=0$ as we have done here is equivalent to requiring that the cosmological constant $\Lambda$ in the action vanishes.}

We can now perturb the action around a generic background $\bar{g}_{\mu\nu}$, i.e. $g_{\mu\nu} = \bar{g}_{\mu\nu} + h_{\mu\nu}$, 
where we can decompose $h_{\mu\nu}$ as follows
\ba         \label{eq:fulldecompositionofperturbationtensor}
        h_{\mu\nu}=h^\perp_{\mu\nu}+\bar\D_{\mu}A_{\nu}+\bar\D_\nu A_\mu+(\bar \D_{\mu}\bar\D_{\nu}-\4
\bar g_{\mu\nu}\bar \Box)B+ \frac{1}{4} \bar g_{\mu\nu}h,\nonumber\\
\ea
where $h^\perp_{\mu\nu}$ is the transverse and traceless spin-2 excitation, $A_\mu$ is a transverse vector field, and $(B,~h)$ are two scalar 
degrees of freedom which mix~\cite{D'Hoker}. One can show that the vector mode and the double derivative scalar mode vanish on constant curvature backgrounds~\cite{Biswas:2016etb}~\footnote{
During inflation the Hubble parameter is nearly constant, so taking the background curvature to be constant is a very good approximation.}.
We are  left with two relevant modes, the tensor mode $h^\perp_{\mu\nu}$ and $\phi \equiv h - \Box B,$ i.e.
\ba \label{eq:decompositionofhintoscalarandtensor}
        h_{\mu\nu} = h^\perp_{\mu\nu} + \frac{1}{4}g_{\mu\nu} \phi.
\ea

\section{Scalar fluctuations around inflationary background}

By inserting Eq.~(\ref{eq:decompositionofhintoscalarandtensor}) into the
action Eq.~(\ref{eq:occurenceunreducedaction}), we find that the scalar part of the quadratic variation of  the action is~\cite{Biswas:2016etb}
\ba\label{scalar}
        \delta^2 S_0 =  \int d^4 x \sqrt{-g} \frac{\phi}{64}  \left( 3 \Box  + \bar{R} \right)\Big[ 6 \lambda F_1 \Box - M^2_p \Big]\phi,
\ea
where $\bar{R}$ is the background Ricci scalar. As we can see, the Weyl tensor term ${\cal F}_3(\Box)$ has no effect here and so the action Eq.~(\ref{eq:occurenceunreducedaction}) should produce the same scalar perturbations as local
Starobinsky $R+R^2$ inflation, as was shown in \cite{Koshelev:2016xqb}.

It was shown in~\cite{Craps:2014wga,Koshelev:2016xqb} that then the  scalar power spectrum is given by:
\ba \label{eq:scalarpowerspectrumintermsofphi}
        |\delta_\Phi (\textbf{k},\tau)|^2 = \left.\frac{k^2}{16 \pi^2 a^2} \frac{1}{3\lambda F_1 \bar{R}}\right|_{k=aH}.
\ea     
We can calculate the measured power spectrum of the gauge-invariant co-moving curvature perturbation $\mathcal{R}$, where in the regime $\dot{H} \ll H^2$,  $\mathcal{R} \approx - \frac{H^2}{\dot{H}}\Phi$, where dot is defined as derivative with respect to physical time, $t$. 
Because the Weyl term does not contribute, then at the crossing of the Hubble radius, the scalar power spectrum is the same as in~\cite{Craps:2014wga,Koshelev:2016xqb}
\ba \label{eq:nonlocalscalarpowerspectrum}
        P_s=|\delta_\mathcal{R}|^2 \approx \frac{H^6_{k=H a}}{16 \pi^2 \dot{H}^2_{k=Ha}} \frac{1}{3\lambda F_1 \bar{R}},
\ea     
evaluated at $k$=$aH$. In the above we have multiplied by $H^4/\dot{H}^2$ in order to transform from the variation of $\Phi$ to the variation of 
$\mathcal{R}$. We can now recast this in terms of number of e-foldings, i.e. $N$,
defined at the start of inflation to the end of inflation~\cite{Mazumdar:2010sa},
\ba \label{eq:defnofnefolds}
        N = - \frac{1}{2} \frac{H^2}{\dot{H}},
\ea 
\vspace{0.1mm}

With the help of the above equation, and noting that the background Ricci scalar during inflation is $\bar{R}\approx 12H^2$, which gives us~\cite{Koshelev:2016xqb}
\ba \label{eq:powerspectrumintermsofN}
        P_s=|\delta_\mathcal{R}|^2 \approx \frac{N^2}{24 \pi^2 } \frac{1}{6\lambda F_1 },
\ea
which reduces to the value for Starobinsky inflation~\cite{Huang:2013hsb}
when we take the appropriate value for $\lambda F_1$, i.e $\lambda F_1=1/(6M_s^2)$, in which case our action reduces to that of Starobinsky. 

This is what we expected, because there is no contribution here from the Weyl tensor term.
Then using Eq.~(\ref{eq:powerspectrumintermsofN}), the scalar spectral tilt is given by~\footnote{Normally $n_s$ is given in terms of $k$ as $n_s=\frac{d (\ln P_s)}{d(\ln k)}$, 
but we write it in terms of $N=\ln(aH)$ using $\frac{d (\ln P_s)}{d(\ln k)}=\frac{1}{P_s} \frac{dP_s}{dN} \frac{dN}{d(\ln k)}$
and then noting that $ \frac{dN}{d(\ln k)} = \frac{d(\ln aH)}{d(\ln aH)}=1.$}
\ba \label{eq:scalarspectralindex}
        n_s=\frac{1}{P_s} \frac{dP_s}{dN} = 1 - \frac{2}{N},
\ea                
where $N$ was defined in Eq.~(\ref{eq:defnofnefolds}). This is the same result as for Starobinsky inflation~\cite{Mukhanov:1990me}. 
\section{Tensor perturbations around inflationary background}

Tensor modes do not couple to the inflaton field in standard inflation. The sub-Hubble tensor modes describe free gravitational waves inside the Hubble patch, which during inflation are carried outside the Hubble patch so that on super-Hubble scales they are locked in. 
We will now calculate the tensor perturbations for our action. While the addition of the Weyl term does not affect the scalar perturbations,
it does affect the tensor perturbations. 
When we insert Eq.~(\ref{eq:decompositionofhintoscalarandtensor})
into the action Eq.~(\ref{eq:occurenceunreducedaction}), then the tensor part of the
quadratic variation of the action is~\cite{Biswas:2016etb}~\footnote{Note that compared to~\cite{Biswas:2016etb}, we took ${\cal F}_2(\Box)=0$ and
$f_{1_0}=F_1$.}
\begin{widetext}
\ba \label{eq:tensorperturbationaction}
\begin{split}
        \delta^2 S_\perp =\frac{\lambda}{4} \int d^4 x \sqrt{-\bar{g}}
 h^\perp_{\mu\nu} \left(\Box - \frac{\bar{R}}{6} \right) F_1 \bar{R} 
        \left[1 + \frac{1}{ F_1 \bar{R}}\left( \Box - \frac{\bar{R}}{3}
\right) {\cal F}_3 \left( \Box + \frac{\bar{R}}{3} \right) \right]
        h^{\perp\mu\nu}
\end{split}
\ea
\end{widetext}
The result Eq.~(\ref{eq:tensorperturbationaction}) 
has the standard pole of the propagator at $\Box=\frac{\bar{R}}{6}$. Eq.~(\ref{eq:tensorperturbationaction}) is simply the result for an action of type 
$R + \lambda F_1 R^2$,  
multiplied by the extra factor in the square brackets. In order to not introduce any ghosts into the propagator, we require that there are no extra poles resulting from this term in square brackets. 
The obvious choice is the exponential of an entire function, which by definition has no roots.
We therefore define~\cite{Koshelev:2016xqb}
\ba \label{eq:definitionofexpomega}
        P(\Box) \equiv 1 + \frac{1}{F_1 \bar{R}} \left(\Box - \frac{\bar{R}}{3}
\right) {\cal F}_3\left( \Box + \frac{\bar{R}}{3} \right),~~~
\ea  
where $P(\Box)$ is the exponential of an entire function. We now look at the precise form of $P(\Box)$. 
If we take the simplest choice, $P(\Box) = e^{\omega(\Box)}$, then we find that 
\ba 
        {\cal F}_3(\Box) = F_1 \bar{R} \frac{e^{\omega(\Box-\bar{R}/3)}-1}{\Box - \frac{2}{3} \bar{R}},
\ea
However, this gives us a pole in $F_3(\Box)$ at $\Box=2\bar{R}/3$.~\footnote{A pole in $F_3(\Box)$ is not necessarily disastrous for the theory, 
because the propagators are still well defined, but having a function $F_3(\Box)$ which is analytic shows that the theory is well\\ constructed.}
The simplest choice which
avoids this pole is~\cite{Koshelev:2016xqb}
\ba \label{eq:choiceoffthree}
        {\cal F}_3(\Box) = F_1 \bar{R} \frac{e^{H(\Box - \frac{2}{3} \bar{R})}-1}{\Box - \frac{2}{3} \bar{R}},
\ea
where $H(\Box)$ is an entire function. Then combining Eq.~(\ref{eq:definitionofexpomega}) with this choice means that
\ba \label{eq:expofomegaintermsofexpH}
        P(\Box)=e^{H(\Box-\bar{R}/3)}.
\ea

The extra exponential factor is always positive and will be very important when we look at the scalar-tensor ratio and will allow us to put a constraint on the scale of non-locality. 
The tensor power spectrum for the action Eq.~(\ref{eq:occurenceunreducedaction}) is therefore multiplied by $P(\Box)$ evaluated at $\Box=\bar{R}/6$, the root of Eq.~(\ref{eq:tensorperturbationaction}). Therefore the power spectrum 
becomes~\cite{Koshelev:2016xqb}
\ba \label{eq:occurencetensorpowerspectrum}
        |\delta_h|^2 = \frac{H^2}{2 \pi^2 \lambda F_1 \bar{R}} e^{{H(-\bar{R}/6)}}.
\ea
and the ratio between the tensor and scalar power spectrums can be given by:
\ba
        r&=& \frac{2|\delta_h|^2}{|\delta_R|^2}\nonumber\\
        &=& 48 H^2 e^{{H(-\bar{R}/6)}}
        \frac{\dot{H}^2}{H^4},
\ea
(where the factor of 2 accounts for the two polarisations of the tensor modes). We can write this tensor-scalar ratio using the definition of the number of e-foldings $N$ Eq.~(\ref{eq:defnofnefolds}) as    
\ba \label{eq:tensorscalarratio}
        r = \frac{12}{N^2} e^{{H(-\bar{R}/6)}}.
\ea                     
When we compare this to the ratio given by Starobinsky inflation, where
\ba \label{eq:starobratio}
        r = \frac{12}{N^2},
\ea
we see that there is an extra modulating exponential factor $e^{H(-\bar{R}/6)}$, 
which was defined in Eq.~(\ref{eq:definitionofexpomega}), due to the addition of non-local gravity.

\section{Scalar to tensor ratio, $r$, and Constraining the scale of non-locality}

From the 2015 Planck data given in~\cite{Ade:2015xua}, the bound on tensor to scalar ratio is given by: $r<0.07$, and so from Eq.~(\ref{eq:tensorscalarratio}), we find that 
\ba \label{eq:constraintexponential}
      \frac{12}{N^2}e^{H(-\bar{R}/6)}<0.07.  
\ea 
When we take the logarithm of Eq.~(\ref{eq:constraintexponential}), we obtain the constraint 
\ba \label{eq:constraintonfcfromr}      
        H(-\bar{R}/6) &<&  2\log(N) - 5.14\,,
\ea    
and during inflation, we obtain:
\ba \label{eq:valueofbackgroundR}
         \bar{R}\approx 12 H^2 \sim \frac{9.02 \times 10^{32}}{N^2} \text{GeV}^2.
\ea        
Therefore  Eq.~(\ref{eq:constraintonfcfromr}) becomes 
\ba\label{fc}
        H\left(-\frac{9.02 \times 10^{32}\text{GeV}^2}{6M^2N^2} \right) &<&  2\log( N) -5.14.
\ea 

Of course, as far as the form of $H(\Box-\bar{R}/3)$ in Eq.~(\ref{eq:expofomegaintermsofexpH}) is concerned, it could be any entire function, 
as long as it retains that in momentum space, $H(-k^2-\bar{R}/3)\to \infty$, when we take the UV limit $k^2\to \infty$. 
This allows the UV propagator to be exponentially suppressed.
This requirement implies that $H$ must have terms of the form $(-1)^n(\Box-\bar{R}/3)^n$
due to the fact that $\Box$ becomes $-k^2$ in momentum space. One can take some 
simple polynomial functions to see what the effect will be on the scalar-tensor ratio.
Note that when we evaluate $r$ at $\Box=\bar{R}/6$, these terms become $(-1)^n(\bar{R}/6-\bar{R}/3)^n=(\bar{R}/6)^n$.
As this is strictly positive, this will gives us an increased value of $r$ compared to that given for Starobinsky inflation 
 in Eq.~(\ref{eq:starobratio}), which means we can then constrain $M$ using the upper bound on $r$ from the Planck data. We will now take

\begin{itemize}

\item 
$H(-\frac{\bar{R}}{6})$=$(-1)^n(\Box-\bar{R}/3)M^2)^n|_{\Box=\bar{R}/6}$=$(\bar{R}/6M^2)^n$   We wish to seek a bound on $M$, which translates Eq.~(\ref{fc}) 
into~\footnote{Note that $50<N<60$ and 2$\log(N)\geq 2\log(50)=7.9$ so therefore $ 2\log(N) -5.14\geq 0$.}:
\ba
        M > \frac{\sqrt{\frac{3}{2}}}{N}  \left[(2\log( N) -5.14)\right]^{-1/2n}  \times 10^{16} \text{GeV}.
\ea

If we take $H(-\bar{R}/6)=-(\Box-\bar{R}/3)/M^2|_{\Box=\bar{R}/6}$, as well as $r<0.07$ and $N$=$60$ e-foldings, then 
\ba
        M> 1.17 \times 10^{14} \text{GeV}.
\ea        
If we take $H(-\bar{R}/6)=(\Box-\bar{R}/3)^2/M^4|_{\Box=\bar{R}/6}$, $r<0.07$ and $N$=$60$ e-foldings, then 
\ba
        M> 1.55 \times 10^{14} \text{GeV}.
\ea        
If we take $H(-\bar{R}/6)$=$(\Box-\bar{R}/3)^{2n}/M^{2n}|_{\Box=\bar{R}/6}$, we obtain:
\ba
        M> 2.04 \times \left(0.573\right)^{1/n} \times 10^{14} \text{GeV}.
\ea
\item $H(-\frac{\bar{R}}{6})$=$-(\Box$-$\frac{\bar{R}}{3})/M^2$+$(-1)^n(\Box$-$\frac{\bar{R}}{3})^a/M^{2a}|_{\Box=\bar{R}/6}$,
Let us illustrate the more general situation, when $H(\Box)$ is a binomial.
There is an extra degree of freedom here because the coefficients in front of the two terms could be different, but to keep things simple we have assumed that they are the same.
In which case, using that $r<0.07$ and $N=60$, our constraint Eq.~(\ref{eq:constraintonfcfromr}) becomes
\ba \label{eq:constraintforbinomial}
        \frac{9.02 \times 10^{32} \text{Gev}}{6M^2 60^2}+ \left(\frac{9.02 \times 10^{32} \text{Gev}}{6M^2 60^2}\right)^a < 3.049~~~
\ea
This gives us a lower bound on $M$ of  $1.78 \times 10^{14}$ GeV for $a=2$, rising to $1.96 \times 10^{14}$ GeV for $a=16$ 
and $2.03 \times 10^{14}$ GeV for $a=64$.

\item $H(-\frac{\bar{R}}{6})= \sum^\infty_{a=1}(-1)^a(\Box-\bar{R}/3)^{a}/M^{2a}|_{\Box=\bar{R}/6}$: 
Finally we take the case where $H$ is a sum of $\Box$ over all orders, again assuming that the coefficients of the terms
are the same. Then our Eq.~(\ref{eq:constraintonfcfromr}) becomes 
\ba
        \sum^\infty_{a=1}\left(\frac{9.02
\times 10^{32} \text{Gev}}{6 \times 60^2M^2}\right)^{2a}< 30.49
\ea               Numerically, this gives us a constraint of
$M>2.35\times 10^{14}$GeV.

\item
So far we have taken only forms of $P(\Box)$ which give a value larger than 1 for $P(\Box=\bar{R}/6)$, but it is also possible
for us to choose a form of $P(\Box)$ such that $P(\bar{R}/6)$ is less than 1.\footnote{For example by taking $P(\Box)=\exp[\Box(\Box-\bar{R}/3)]$.
This still ensures that we have no pole in $F_3(\Box)$, but gives us a negative value for the argument of the exponential when we evaluate $P(\Box)$ at $\Box=\bar{R}/6$.}
Using Eq.~(\ref{eq:tensorscalarratio}), a negative value of for the argument of the exponential gives a lower tensor-scalar ratio
than Starobinsky inflation in Eq.~(\ref{eq:starobratio}), and because we do not currently have a lower bound on $r$, this does not provide us with any constraint on $M$. 
If future data shows that $r$ is lower than that predicted by Starobinsky,
then this could prove a useful model.\footnote{We could also take $P(\Box)=e^{(\Box-\bar{R}/6)(\Box-\bar{R}/3)}$. When we evaluate this at $\Box = \bar{R}/6$, then this gives us the
same prediction as Starobinsky.}
\end{itemize}
\begin{figure}[]
\includegraphics[width=8cm]{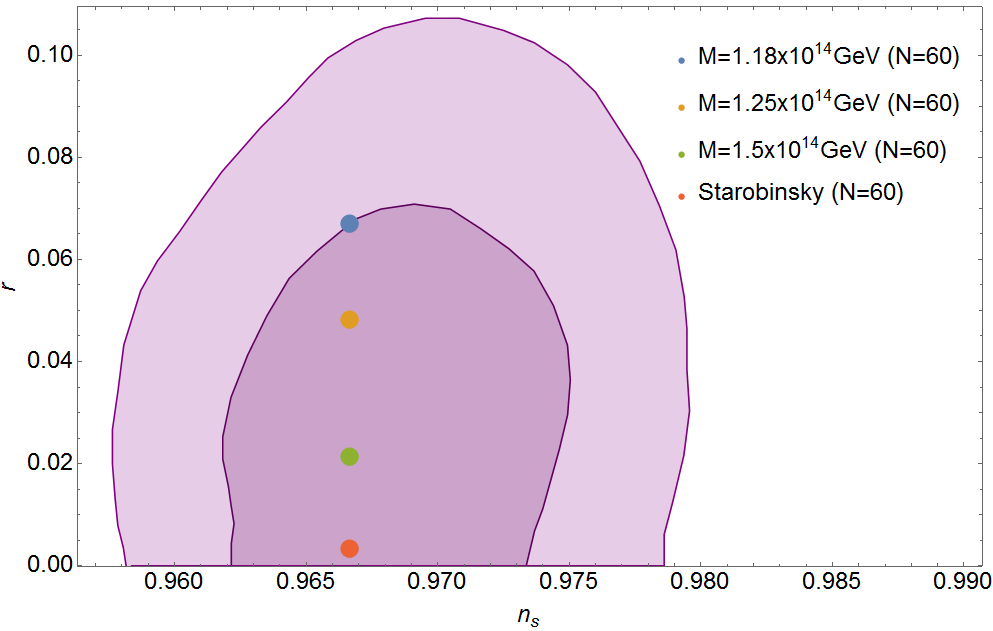}
\caption{A plot of the tensor-scalar ratio $r$ vs the spectral index $n_s$ for different values of $M$ 
where we have taken $H(-\bar{R}/6)=-(\Box-\bar{R}/3)/M^2|_{\Box=\bar{R}/6}$ using Eq.~(\ref{eq:rintermsofns}).  
We have taken $N=60$ and also plotted the 2015 Planck data.}
\label{fig:plotofrvsnsintermsofH}
\end{figure}
\section{$r$ vs. $n_s $ plot for UV complete Starobinsky model of inflation}
\begin{table}[H]
\begin{center}
  \begin{tabular}{ |l   r| }
    \hline
    $M$ &  Tensor-scalar ratio \\ \hline
    $1\times10^{14}$GeV & 0.22  \\ 
    $1.5\times10^{14}$GeV & 0.021  \\ 
    $2.0\times10^{14}$GeV & 0.0094  \\ 
    $M_p=2.435\times 10^{18}$GeV & 0.0033 \\
    $\infty$ (Starobinsky) &  0.0033 \\
   \hline
  \end{tabular}
\caption{A table of the tensor-scalar ratio for various values of the scale of non-locality $M$,
using Eq.~(\ref{eq:valueofbackgroundR}) and Eq.~(\ref{eq:rintermsofns}) and taking $H(-\bar{R}/6)=-(\Box-\bar{R}/3)/M^2|_{\Box=\bar{R}/6}$. 
In the fifth line, we have taken the limit $M\to \infty$, 
which means that our action reduces to that of Starobinsky. 
The tensor-scalar ratio for $M=M_p$ is a factor of $\exp(7.04\times 10^{-9})$
bigger than for Starobinsky inflation, i.e. when $M\to\infty$.
This factor is equal to 1 to nine significant figures, i.e. for $M=M_p$,
the action Eq.~(\ref{eq:occurenceunreducedaction}) is effectively indistinguishable from Starobinsky inflation.}
\end{center}
\end{table}
If we want to plot $r$ against $n_s$, then we should note using Eq.~(\ref{eq:scalarspectralindex}), Eq.~(\ref{eq:tensorscalarratio})  
that for non-local gravity  
\ba \label{eq:rintermsofns}
        r &=& \frac{12}{N^2} e^{H(\bar{R}/6)}\nonumber\\
        &=& 3(1-n_s)^2 e^{H (-\bar{R}/6)}.
\ea        
We can plot this for different forms of $H(\Box)$ and compare with the 2015 Planck data at the 68\% and 95\% confidence level from~\cite{Ade:2015xua} for $N$=60. 
If we take $H(\Box)= -\Box/M^2$, then in Fig.~[\ref{fig:plotofrvsnsintermsofH}]
this gives us a constraint on $M$ of 
\ba
        M>  1.18\times 10^{14}\text{GeV}.
\ea

\section{Conclusion}
In recent years, an 
\textit{infinite derivative, ghost-free, quadratic curvature} action of gravity has
been shown to solve the singularity problem and give a universal prediction
for the Newtonian potential at large distances.

Using the scalar spectral index and the tensor-scalar ratio together with
the latest Planck data,  
we have found a constraint on the value of the scale of non-locality for
various cases of infinite derivative gravity (IDG). 

Using the simplest case of IDG which avoids poles throughout the theory,
this provides us with a much stronger constraint than before, of $M$$>$$1.18
\times 10^{14}$ GeV (around $10^{-4} M_P$) 
using cosmological data whereas previously our best constraint from below
using the full theory was $10^{-2}$ eV using data from laboratory experiments.
Using a reduced version of the action, the constraints $M>10^8$~GeV~\cite{Biswas:2013dry} and $M\sim 10^{15}$~GeV~\cite{Biswas:2006bs} were found using cosmological data by other authors, which is consistent with our result. 

We have also looked at different classes of the theory, with different versions
of the entire function, and obtained modified constraints within these versions.
In principle we can take a different argument of the function, which will
give us different values of the tensor-scalar ratio.  

Our result allows us to compare experimentally IDG with Starobinsky
inflation, and can provide an explanation for possible differences between the predictions of Starobinsky inflation and cosmological data.
With further data on the scalar-tensor ratio we will be able to constrain
the scale of non-locality even further.

\section*{Acknowledgements}
We would like to thank Alexey Koshelev and Anupam Mazumdar for valuable discussions and guidance,
without which this paper would not have been possible.


\end{document}